%% file: diphoton.tex
\documentclass[preprint,twocolumn,5p]{elsarticle}

\usepackage{graphicx}
\usepackage{xspace}
\usepackage{units}
\usepackage{amssymb}
\usepackage[hyperindex,breaklinks]{hyperref}

\biboptions{compress}

\newcommand{\CL}{CL\xspace}

\newcommand{\ptm}{\ensuremath{{\bf p}_{\mathrm{T}}^{\rm miss}}\xspace}
\newcommand{\MET}{\ensuremath{E_{\mathrm{T}}^{\mathrm{miss}}}\xspace}
\newcommand{\met}{\MET}
\newcommand{\dphi}{\ensuremath{\Delta \phi(\gamma,\MET)}\xspace}
\newcommand{\dphim}{\ensuremath{\Delta \phi_{\mathrm{min}}(\gamma,\MET)}\xspace}
\newcommand{\HT}{\ensuremath{H_{\mathrm{T}}}\xspace} 
\newcommand{\Section}{Section}
\newcommand{\Fig}{Fig.\xspace}
\newcommand{\Ref}{Ref.\xspace}
\newcommand{\Tab}{Table\xspace}

\newcommand{\mass}[1]{\ensuremath{m_{#1}}}
\newcommand{\ET}{\ensuremath{E_{\mathrm{T}}}\xspace}
\newcommand{\pt}{\ensuremath{p_{\mathrm{T}}}\xspace}

\newcommand{\antibar}[1]{\ensuremath{#1\bar{#1}}}
\newcommand{\ttbar}{\antibar{t}\xspace}

\newcommand{\myW}{\ensuremath{W(\to e\nu)+\mathrm{jets}/\gamma}\xspace}
\newcommand{\myttbar}{\ensuremath{\ttbar(\to e\nu)+\mathrm{jets}}\xspace}
\newcommand{\Zgg}{\ensuremath{Z(\to \nu\bar{\nu})+\gamma\gamma}\xspace}
\newcommand{\Wgg}{\ensuremath{W(\to \ell\nu)+\gamma\gamma}\xspace}

\newcommand{\ifb}{\mbox{fb$^{-1}$}}

\newcommand{\Nfive}{\ensuremath{N_{5}}\xspace}

\newcommand{\QCDg}{\ensuremath{\mathrm{QCD}}\xspace}
\newcommand{\neutralino}{\ensuremath{\tilde{\chi}^{0}_{1}}\xspace}
\newcommand{\neutralinotwo}{\ensuremath{\tilde{\chi}^{0}_{2}}\xspace}
\newcommand{\chargino}{\ensuremath{\tilde{\chi}^{\pm}_{1}}\xspace}
\newcommand{\chiplus}{\ensuremath{\tilde{\chi}^{+}_{1}}\xspace}
\newcommand{\chiminus}{\ensuremath{\tilde{\chi}^{-}_{1}}\xspace}
\newcommand{\gravitino}{\ensuremath{\tilde{G}}\xspace}
\newcommand{\gluino}{\ensuremath{\tilde{g}}\xspace}
\newcommand{\Pythia}{{\tt PYTHIA}\xspace}
\newcommand{\Herwigpp}{{\tt Herwig++}\xspace}
\newcommand{\Madgraph}{{\tt MadGraph}\xspace}
\newcommand{\Suspect}{{\tt SUSPECT}\xspace}
\newcommand{\Sdecay}{{\tt SDECAY}\xspace}
\newcommand{\Isajet}{{\tt ISAJET}\xspace}

\newcommand{\integLumi}{\unit[4.8]{\ifb}\xspace}
\newcommand{\integLumiE}{\unit[$(4.8\pm 0.2)$]{\ifb}\xspace}

\newcommand{\UEDlimitlow}{\unit[1.40]{TeV}\xspace}

\newcommand{\GGMlimitg}{\unit[1.07]{TeV}\xspace}

\newcommand{\GGMlimitsq}{\unit[0.87]{TeV}\xspace}

\newcommand{\SPSlimit}{\unit[196]{TeV}\xspace}

\usepackage{preprintcover} 
\PreprintCoverPaperTitle{Search for Diphoton Events with Large Missing
  Transverse Momentum in \unit[7]{TeV} Proton--Proton Collision Data
  with the ATLAS Detector} 
\PreprintIdNumber{CERN-PH-EP-2012-234} 
\PreprintCoverAbstract{ A search for diphoton events with large
  missing transverse momentum has been performed using proton--proton
  collision data at $\sqrt{s} = \unit[7]{TeV}$ recorded with the ATLAS
  detector, corresponding to an integrated luminosity of \integLumi.
  No excess of events was observed above the Standard Model prediction
  and model-dependent \unit[95]{\%}~confidence level exclusion limits
  are set. In the context of a generalised model of gauge-mediated
  supersymmetry breaking with a bino-like lightest neutralino of mass
  above 50 GeV, gluinos (squarks) below \GGMlimitg (\GGMlimitsq) are
  excluded, while a breaking scale $\Lambda$ below \SPSlimit is
  excluded for a minimal model of gauge-mediated supersymmetry
  breaking.  For a specific model with one universal extra dimension,
  compactification scales $1/R <$ \UEDlimitlow are excluded. These
  limits provide the most stringent tests of these models to date.  }
\PreprintJournalName{Physics Letters B}  

\journal{Physics Letters B}

\begin{document}

\begin{frontmatter}

\title{Search for Diphoton Events with Large Missing Transverse Momentum in \\
\unit[7]{TeV} Proton--Proton Collision Data with the ATLAS Detector
}

\author{The ATLAS Collaboration}

\begin{abstract}
  A search for diphoton events with large missing transverse momentum
  has been performed using proton--proton collision data at $\sqrt{s}
  = \unit[7]{TeV}$ recorded with the ATLAS detector, corresponding to
  an integrated luminosity of \integLumi.  No excess of events was
  observed above the Standard Model prediction and model-dependent
  \unit[95]{\%}~confidence level exclusion limits are set. In the
  context of a generalised model of gauge-mediated supersymmetry
  breaking with a bino-like lightest neutralino of mass above 50 GeV,
  gluinos (squarks) below \GGMlimitg (\GGMlimitsq) are excluded, while
  a breaking scale $\Lambda$ below \SPSlimit is excluded for a minimal
  model of gauge-mediated supersymmetry breaking.  For a specific
  model with one universal extra dimension, compactification scales
  $1/R <$ \UEDlimitlow are excluded. These limits provide the most
  stringent tests of these models to date.
\end{abstract}

\end{frontmatter}

\section{Introduction}
This Letter reports on a search for diphoton ($\gamma\gamma$) events
with large missing transverse momentum (\MET) in \integLumi of
proton--proton ($pp$) collision data at $\sqrt{s}=\unit[7]{TeV}$
recorded with the ATLAS detector at the Large Hadron Collider (LHC)
in 2011, extending and superseding 
a prior study performed with \unit[1]{\ifb}~\cite{Aad:2011zj}.  The
results are interpreted in the context of three models of new physics:
a general model of gauge-mediated supersymmetry breaking
(GGM)~\cite{Meade:2008wd,Buican:2008ws,Ruderman:2011vv}, a minimal
model of gauge-mediated supersymmetry breaking
(SPS8)~\cite{Allanach:2002nj}, and a model with one universal
extra dimension
(UED)~\cite{Appelquist:2000nn,Macesanu:2002ew,Macesanu:2005jx}.

\section{Supersymmetry}
\label{sec:susy}
Supersymmetry~(SUSY)~\cite{Miyazawa:1966,Ramond:1971gb,Golfand:1971iw,Neveu:1971rx,Neveu:1971iv,Gervais:1971ji,Volkov:1973ix,Wess:1973kz,Wess:1974tw} 
introduces a symmetry between fermions and bosons, resulting in a SUSY
partner (sparticle) with identical quantum numbers except a difference
by half a unit of spin for each Standard Model (SM) particle. As none
of these sparticles have been observed, SUSY must be a broken symmetry
if realised in nature.  Assuming $R$-parity
conservation~\cite{Fayet:1976et,Fayet:1977yc,Farrar:1978xj,Fayet:1979sa,Dimopoulos:1981zb},
sparticles are
produced in pairs.  These would then decay through cascades involving
other sparticles until the lightest SUSY particle (LSP), which is stable, is produced.

In gauge-mediated SUSY breaking~(GMSB)
models~\cite{Dine:1981gu,AlvarezGaume:1981wy,Nappi:1982hm,Dine:1993yw, Dine:1994vc,Dine:1995ag} 
the LSP is the gravitino \gravitino. GMSB experimental signatures are
largely determined by the nature of the next-to-lightest SUSY
particle~(NLSP). In this study the NLSP is assumed to be
the lightest neutralino \neutralino.
For studies with the lightest stau
as NLSP, the reader is referred to Refs.~\cite{Aad:2012ag,Aad:2012rt}.
Should the lightest neutralino be a bino 
(the SUSY partner of the SM U(1) gauge boson),
the final decay in the cascade would predominantly be
$\neutralino\to\gamma\gravitino$, with two cascades per event, leading
to final states with $\gamma\gamma+\met$, where $\met$ results from
the undetected gravitinos.

Two different classes of gauge-mediated models, described in more detail below, 
are considered as benchmarks to evaluate the reach of this analysis:
the minimal GMSB model (SPS8)
as an example of a
complete SUSY model with a full particle spectrum and two different
variants of the GGM model as examples of phenomenological models with
reduced particle content.

In the SPS8 model, the only free parameter is the SUSY-breaking mass scale $\Lambda$
that establishes the nature of the observable phenomena exhibited by
the low-energy sector. 
The other model
parameters are fixed to the following values: the messenger mass
$M_{\mathrm{mess}}=2\Lambda$, the number of SU(5) messengers
$\Nfive=1$, the ratio of the vacuum expectation values of the two
Higgs doublets $\tan\beta=15$, and the Higgs sector mixing parameter
$\mu>0$. The bino NLSP is assumed to decay promptly ($c\tau_{\mathrm{NLSP}}
< \unit[0.1]{mm}$).
For $\Lambda \simeq \unit[200]{TeV}$, the direct production of gaugino pairs
such as \neutralinotwo\chargino or \chiplus\chiminus pairs is
expected to dominate at a LHC centre-of-mass energy of
$\sqrt{s}=\unit[7]{TeV}$.
The contribution from gluino and/or squark pairs is
below \unit[10]{\%} of the production cross section due to their high
masses.
The sparticle pair produced in the collision decays via cascades into
two photons and two gravitinos. Further SM particles such as gluons, quarks, leptons
and gauge bosons may be produced in the cascade decays.
The current best limit on $\Lambda$ in this model is
\unit[145]{TeV}~\cite{Aad:2011zj}.

Two different configurations of the GGM SUSY model
are considered in this study, for which the neutralino
NLSP, chosen to be the bino, and either the
gluino or the squark masses are treated as free parameters.
For the squark--bino GGM model all squark masses are treated as
degenerate except the right-handed up-type squarks whose mass is
decoupled (set to inaccessibly large values). 
For the gluino--bino model all squark masses are
decoupled. 
For both configurations
all other sparticle masses are
also decoupled,
leading to a dominant production mode
at $\sqrt{s} = \unit[7]{TeV}$ of a pair of squarks in one case and a
pair of gluinos in the other case. 
These would decay via short cascades into the bino-like neutralino
NLSP. Jets may be produced in the cascades from the gluino and squark
decays. Further model parameters are fixed to $\tan\beta=2$ and
$c\tau_{\mathrm{NLSP}} < \unit[0.1]{mm}$.
The decay into the wino-like neutralino NLSP is possible and was
studied by the CMS Collaboration~\cite{Chatrchyan:2011ah}.

\section{Extra dimensions}
UED models postulate the existence of additional spatial dimensions in
which all SM particles can propagate, leading to the existence of a
series of excitations for each SM particle, known as a Kaluza--Klein
(KK) tower. This analysis considers the case of a single UED, with
compactification radius (size of the extra dimension) $R \approx
\unit[1]{TeV^{-1}}$.
At the LHC, the main UED process would be the production via the
strong interaction of a pair of first-level KK quarks and/or
gluons~\cite{Macesanu:2002db}. These would decay via cascades
involving other KK particles until reaching the lightest KK particle
(LKP), i.e. the first level KK photon $\gamma^*$. SM particles
such as quarks, gluons, leptons and gauge bosons may be produced in the
cascades. If the UED model is embedded in a larger space with $N$
additional eV$^{-1}$-sized dimensions accessible only to
gravity~\cite{DeRujula:2000he}, with a $(4 + N)$-dimensional Planck
scale ($M_D$) of a few \unit{TeV}, the LKP would decay gravitationally
via $\gamma ^* \rightarrow \gamma + G$. $G$ represents a tower of
eV-spaced graviton states, leading to a graviton mass between $0$ and
$1/R$.
With two decay chains per event, the final state would contain $\gamma
\gamma + \MET$, where \MET results from the escaping gravitons.
Up to $1/R \sim\unit[1]{TeV}$, the branching ratio to the diphoton and
\MET final state is close to \unit[100]{\%}. As $1/R$ increases, the
gravitational decay widths become more important for all KK particles
and the branching ratio into photons decreases, e.g. to \unit[50]{\%}
for $1/R=\unit[1.5]{TeV}$~\cite{Macesanu:2002ew}.

The UED model considered here is defined by specifying $R$ and
$\Lambda$, the ultraviolet cut-off used in the calculation of
radiative corrections to the KK masses. This analysis sets $\Lambda $
such that $\Lambda R = 20$~\cite{Cheng:2002iz}. The $\gamma^*$ mass is insensitive to
$\Lambda$, while other KK masses typically change by a few per cent
when varying $\Lambda R$ in the range $10-30$. For $1/R =
\unit[1.4]{TeV}$, the masses of the first-level KK photon, quark and
gluon are $1.40$ TeV, $1.62$ TeV and \unit[1.71]{TeV},
respectively~\cite{ElKacimi:2009zj}.

\section{Simulated samples}

For the GGM model, the SUSY mass spectra were calculated using
\Suspect 2.41~\cite{Djouadi:2002ze} and
\Sdecay~1.3~\cite{Muhlleitner:2003vg}; for the SPS8 model, the SUSY
mass spectra were calculated using \Isajet~7.80~\cite{isajet}. 
The Monte Carlo~(MC) SUSY signal samples were produced using
\Herwigpp~2.5.1~\cite{Bahr:2008pv} with {\tt MRST2007
  LO${}^{*}$}~\cite{Sherstnev:2007nd} parton distribution
functions~(PDFs).
Signal cross sections were calculated to next-to-leading order~(NLO) in the strong coupling constant, 
including the resummation of soft gluon emission at next-to-leading-logarithmic accuracy 
(NLO+NLL)~\cite{Beenakker:1996ch,Kulesza:2008jb,Kulesza:2009kq,Beenakker:2009ha,Beenakker:2011fu}. 
The nominal cross sections and the uncertainties were taken from an envelope of cross-section predictions 
using different PDF sets and factorisation and renormalisation scales, as described in 
\Ref~\cite{Kramer:2012bx}.
In the case of the UED model, cross sections were estimated and MC signal samples generated using
the UED model as implemented at leading order (LO) in
\Pythia~6.423~\cite{pythia,ElKacimi:2009zj} with {\tt
  MRST2007 LO${}^{*}$}  PDFs.

The ``irreducible'' background from \Wgg and \Zgg production was
simulated at LO using \Madgraph~4~\cite{Alwall:2007st} with the {\tt
  CTEQ6L1}~\cite{cteq6m} PDFs. Parton showering and fragmentation were
simulated with \Pythia. NLO cross sections and scale uncertainties
were implemented via multiplicative constants ($K$-factors)
that relate the NLO and LO cross sections. These have been calculated
for several restricted regions of the overall phase space
of the \Zgg and \Wgg processes~\cite{Bozzi:2011en,Bozzi:2011wwa},
and are estimated to be
$2.0 \pm 0.3$ and $3 \pm 3$ for the \Zgg and \Wgg contributions
to the signal regions of this analysis, respectively.
As described below, all other background sources are estimated
through the use of control samples derived from data.

All samples were processed through the {\tt GEANT4}-based
simulation of the ATLAS detector~\cite{geant4,Aad:2010wq}. 
The variation of the number of $pp$ interactions per bunch crossing
(pile-up) as a function of the instantaneous luminosity is taken into
account by overlaying simulated minimum bias events
according to the observed distribution of the number of pile-up
interactions in data, with an average of $\sim 10$ interactions.

\section{ATLAS detector}
The ATLAS detector~\cite{Aad:2008zzm} is a multi-purpose apparatus
with a forward-backward symmetric cylindrical geometry and nearly
4$\pi$ solid angle coverage.  Closest to the beamline are tracking
devices comprising layers of silicon-based pixel and strip detectors
covering $\left|\eta\right|<2.5$\footnote{ATLAS uses a right-handed
  coordinate system with its origin at the nominal interaction point
  (IP) in the centre of the detector and the $z$-axis along the beam
  pipe. The $x$-axis points from the IP to the centre of the LHC ring,
  and the $y$-axis points upward. Cylindrical coordinates $(r,\phi)$
  are used in the transverse plane, $\phi$ being the azimuthal angle
  around the beam pipe. The pseudorapidity is defined in terms of the
  polar angle $\theta$ as $\eta=-\ln\tan(\theta/2)$.} and straw-tube
detectors covering $\left|\eta\right|<2.0$, located inside a thin
superconducting solenoid that provides a \unit[2]{T} magnetic field.
Outside the solenoid, fine-granularity lead/liquid-argon 
electromagnetic (EM) calorimeters provide coverage for
$\left|\eta\right| < 3.2$ to measure the energy and position of
electrons and photons.
A presampler, covering $\left|\eta\right| < 1.8$, is used to correct
for energy lost upstream of the EM calorimeter.  An
iron/scintillating-tile hadronic calorimeter covers the region $|\eta|
< 1.7$, while a copper/liquid-argon medium is used for hadronic
calorimeters in the end-cap region $1.5 < |\eta| < 3.2$. In the
forward region $3.2 < |\eta| < 4.9$ liquid-argon calorimeters with
copper and tungsten absorbers measure the electromagnetic and hadronic
energy.  A muon spectrometer consisting of three superconducting
toroidal magnet systems each comprising eight toroidal coils, 
tracking chambers, and detectors for
triggering surrounds the calorimeter system.

\section{Reconstruction of candidates and observables}
The reconstruction of converted and unconverted photons and of
electrons is described in Refs.~\cite{Aad:2010sp} and
\cite{Aad:2011mk}, respectively.
Photon candidates were required to be within $\left|\eta\right| <
1.81$, 
and to be outside the
transition region $1.37 < \left|\eta\right| < 1.52$ between the barrel
and end-cap calorimeters.  
Identified on the basis of the characteristics of the longitudinal and transverse shower
development in the EM calorimeter,
the analysis made use of both ``loose'' and
``tight'' photons~\cite{Aad:2010sp}. 
In the case that an EM calorimeter deposition was identified as both 
a photon and an electron, the photon candidate was discarded
and the electron candidate retained.
In addition, converted photons were re-classified
as electrons if one or more candidate conversion tracks included
at least one hit from the pixel layers.
Giving preference to the electron selection 
in this way reduced the electron-to-photon fake
rate by $50$--$60$\% (depending on the value of $\eta$) relative 
to that of the prior \unit[1]{\ifb} analysis~\cite{Aad:2011zj},
while preserving over $70$\% of the signal efficiency.
Finally, an ``isolation'' requirement was imposed.
After correcting for contributions
from pile-up and the deposition ascribed to
the photon itself, photon candidates were removed if more than $\unit[5]{GeV}$ of 
transverse energy was observed in a cone of $\sqrt{(\Delta \eta)^2 + (\Delta \phi)^2} < 0.2$ surrounding the
energy deposition in the calorimeter associated with the photon.

The measurement of the two-dimensional transverse momentum 
vector \ptm (and its magnitude \met) was based on energy deposits in
calorimeter cells inside three-dimensional clusters with $|\eta | <
4.9$ and was corrected for contributions from muons, if
any~\cite{Aad:2012re}. The cluster energy was calibrated to correct for the 
different response to electromagnetically- and hadronically-induced showers,
energy loss in dead
material, and out-of-cluster energy.
The contribution from identified muons was accounted for by adding
in the energy derived from the properties of 
reconstructed muon tracks.

Jets were reconstructed using the anti-$k_t$ jet
algorithm~\cite{Cacciari:2008gp} with 
radius parameter $R=0.4$. They were required to
have $\pt > \unit[20]{GeV}$ and $|\eta| < 2.8$~\cite{Aad:2011he}.

Two additional observables of use in discriminating SM backgrounds
from potential GMSB and UED signals were defined. The total visible transverse energy 
\HT was calculated as the sum of the magnitude of
the transverse momenta of the two selected photons and any
additional leptons and jets in the event. The photon--\met separation
\dphi was defined as the azimuthal
angle between the missing transverse momentum vector and either of the two
selected photons, with \dphim the minimum value of \dphi of the two
selected photons.

\section{Data analysis}

\begin{figure}[tb]
  \begin{center}
    \includegraphics[width=0.48\textwidth]{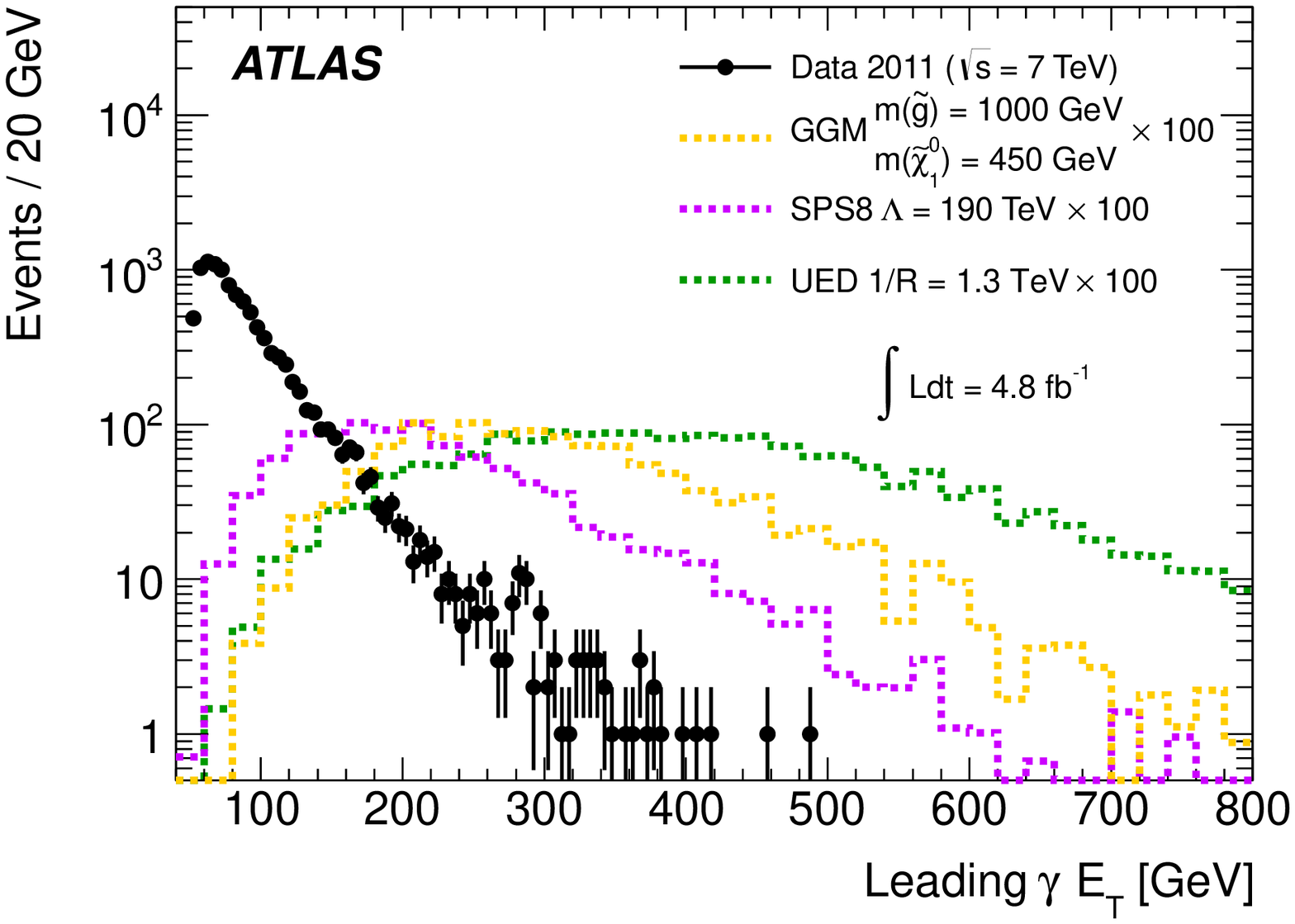}
  \end{center}
  \caption{The \ET spectrum of the leading photon in the
    $\gamma\gamma$ candidate events in the data (points, statistical
    uncertainty only) together with the spectra from simulated GGM
    ($\mass{\gluino} = \unit[1000]{GeV},\mass{\neutralino} = \unit[450]{GeV}$),
    SPS8 ($\Lambda = \unit[190]{TeV}$), and
    UED ($1/R = \unit[1.3]{TeV}$) samples after the diphoton
    requirement. The signal samples are scaled by a factor of 100 for
    clarity.
    \label{fig:PhoPt}
  }
\end{figure}

\begin{figure}[tb]
  \begin{center}
    \includegraphics[width=0.48\textwidth]{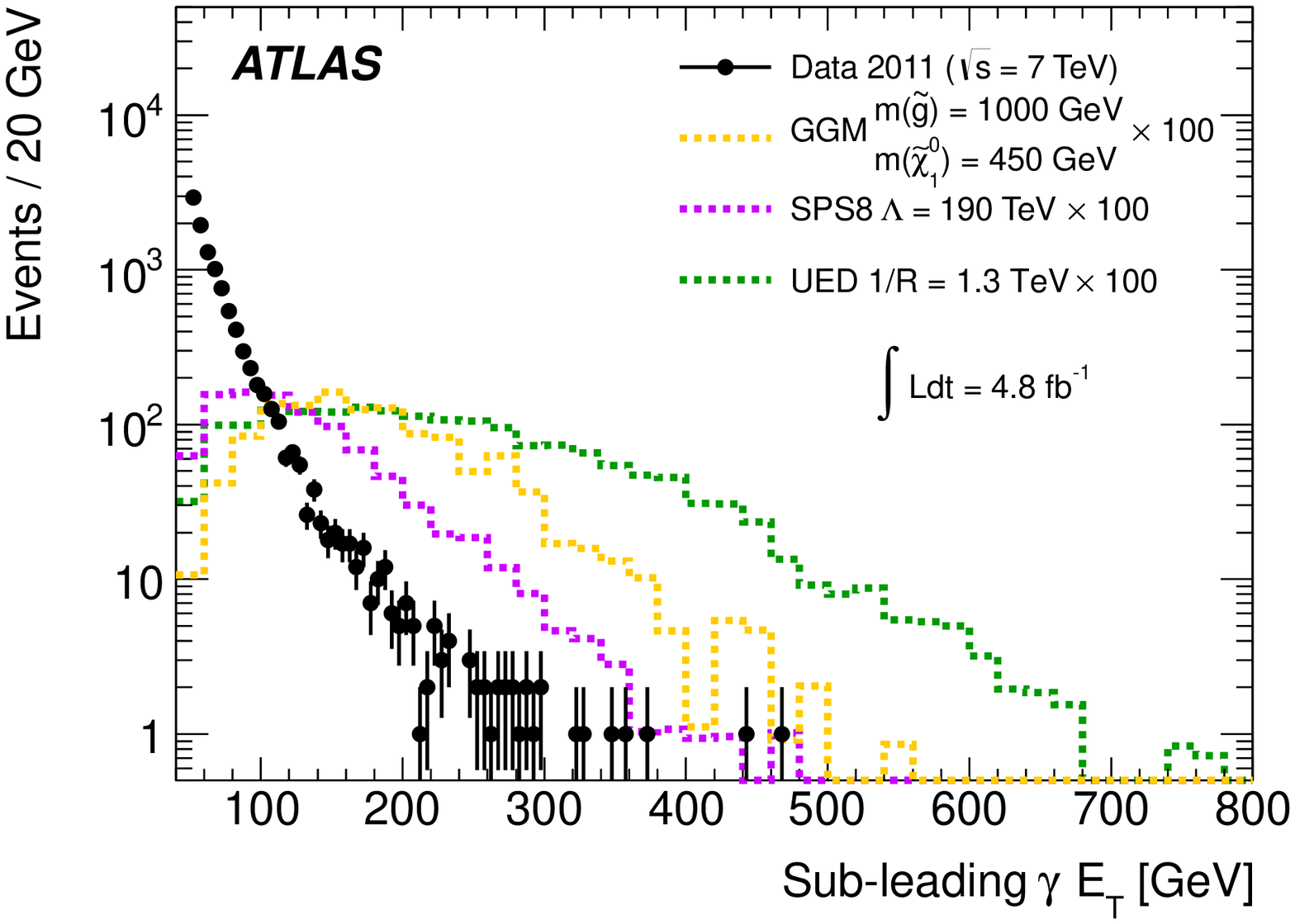}
  \end{center}
  \caption{The \ET spectrum of the sub-leading photon in the
    $\gamma\gamma$ candidate events in the data (points, statistical
    uncertainty only) together with the spectra from simulated GGM
    ($\mass{\gluino} = \unit[1000]{GeV},\mass{\neutralino} = \unit[450]{GeV}$),
    SPS8 ($\Lambda = \unit[190]{TeV}$), and
    UED ($1/R = \unit[1.3]{TeV}$) samples after the diphoton
    requirement. The signal samples are scaled by a factor of 100 for
    clarity.
    \label{fig:Pho2Pt}
  }
\end{figure}

\begin{figure}[tb]
  \begin{center}
    \includegraphics[width=0.48\textwidth]{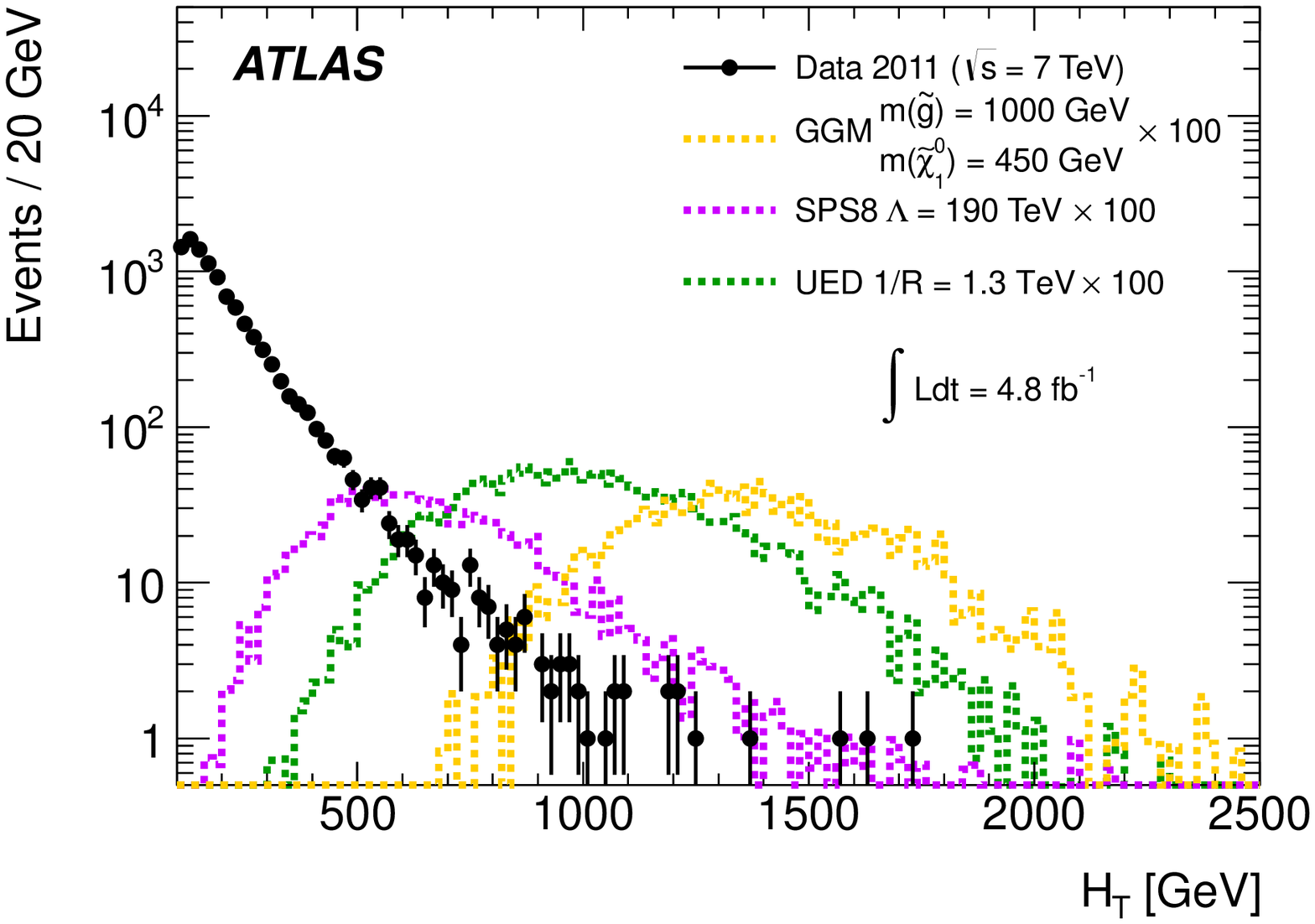}
  \end{center}
  \caption{The \HT spectrum of 
    $\gamma\gamma$ candidate events in the data (points, statistical
    uncertainty only) together with the spectra from simulated GGM
    ($\mass{\gluino} = \unit[1000]{GeV},\mass{\neutralino} = \unit[450]{GeV}$),
    SPS8 ($\Lambda = \unit[190]{TeV}$), and
    UED ($1/R = \unit[1.3]{TeV}$) samples after the diphoton
    requirement. The signal samples are scaled by a factor of 100 for
    clarity.
    \label{fig:HT}
  }
\end{figure}

\begin{figure}[tb]
  \begin{center}
    \includegraphics[width=0.48\textwidth]{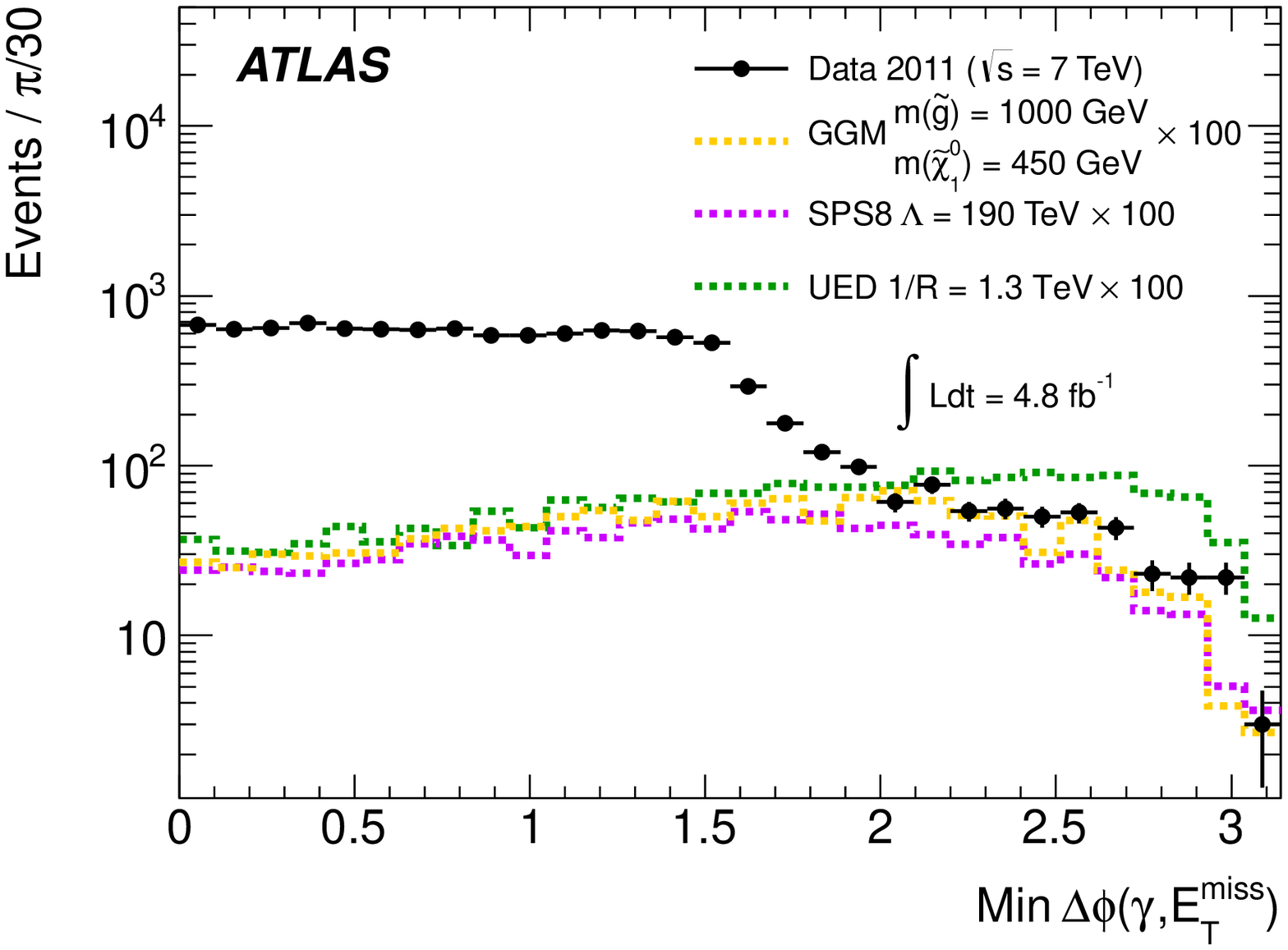}
  \end{center}
  \caption{The minimum \dphi spectrum of 
    $\gamma\gamma$ candidate events in the data (points, statistical
    uncertainty only) together with the spectra from simulated GGM
    ($\mass{\gluino} = \unit[1000]{GeV},\mass{\neutralino} = \unit[450]{GeV}$),
    SPS8 ($\Lambda = \unit[190]{TeV}$), and
    UED ($1/R = \unit[1.3]{TeV}$) samples after the diphoton
    requirement. The signal samples are scaled by a factor of 100 for
    clarity.
    \label{fig:dphi}
  }
\end{figure}

The data sample, corresponding to an integrated luminosity of
\integLumiE~\cite{lumi2011,Aad:2011dr}, was selected by a trigger requiring two loose photon
candidates with $\ET > \unit[20]{GeV}$.
To ensure the event resulted from 
a beam collision, events were required to have at least one vertex 
with five or more associated tracks.
Events were then required to contain at
least two tight photon candidates with $\ET > \unit[50]{GeV}$, which
MC studies suggested would provide the greatest separation between
signal and SM background for a broad range of 
the parameter space
of the new physics
scenarios under consideration in this search.
A total of 10455 isolated $\gamma\gamma$
candidate events passing these selection requirements were observed in the data sample.
The \ET distributions\footnote{
An excess of events 
relative to a smoothly-falling distribution
of the leading-photon spectrum was observed for $\ET \sim \unit[285]{GeV}$.
Searching over the range $\unit[100]{GeV} < \ET < \unit[500]{GeV}$,  
a significance of $\unit[1.9]{\sigma}$ was found using BumpHunter~\cite{Chou:2011}, 
while the local significance was found to be $\unit[3.1]{\sigma}$.
No correlation between the excess and the LHC running period or luminosity was observed.
A comparison of other observables (e.g. diphoton mass, \MET, leading-photon $\eta$, $\Delta\phi(\gamma_1,\gamma_2)$)
between the excess and sideband regions exhibited no appreciable differences.
It was concluded that the observed excess of events is compatible with a statistical fluctuation.
}
of the leading and sub-leading photon for events in
this sample are shown in Figs.~\ref{fig:PhoPt} and~\ref{fig:Pho2Pt}.
Also shown are the \ET
spectra obtained from GGM MC samples for
$\mass{\gluino}=\unit[1000]{GeV}$ and
$\mass{\neutralino}=\unit[450]{GeV}$,
from SPS8 MC samples with $\Lambda = \unit[190]{TeV}$, 
and from UED MC samples for $1/R = \unit[1.3]{TeV}$, representing
model parameters near the expected exclusion limit. 
Figures~\ref{fig:HT}
and~\ref{fig:dphi} show the \HT and \dphim distributions of
selected diphoton events, with those of the same signal models
overlaid.

To maximise the sensitivity of this analysis over a wide
range of model parameters that may lead to different kinematic
properties, three different signal regions (SRs) were defined based on
the observed values of
\met, \HT
and \dphim.
SR~A, optimised for gluino/squark production with a subsequent decay
to a high-mass bino, requires large \met and moderate \HT. 
SR~B, optimised for gluino/squark production with a subsequent decay to a
low-mass bino, requires moderate \MET and large \HT. SR~C, optimised
for the electroweak production of intermediate-mass gaugino pairs that dominates the SPS8
cross section in this regime, requires moderate \MET but makes
no requirement on \HT. In addition, a
requirement of $\dphim > 0.5$ was imposed on events in SR~A and C;
for the low-mass bino targeted by SR~B, the separation between the
photon and gravitino daughters of the bino is too slight to allow for the
efficient separation of signal from background through the use of
this observable. The selection requirements of the three SRs are
summarised in \Tab~\ref{tab:signalregion}. Of the three SRs, SR~A 
provides the greatest sensitivity to the UED model, and is thus the SR 
used to test this model.

\begin{table}[htb]
  \caption{Definition of the three SRs (A, B and C) based on the quantities
    \met, \HT and \dphim.}
  \label{tab:signalregion}
  \centering
  \begin{tabular}{l|ccc}
    \hline
          &   SR A   & SR B &   SR C \\
    \hline
    $\met >$  &  \unit[200]{GeV} &   \unit[100]{GeV} &   \unit[125]{GeV} \\
    $\HT  >$  &  \unit[600]{GeV} &  \unit[1100]{GeV} &  - \\
    $\dphim >$ &  0.5             &   -               & 0.5 \\
    \hline
  \end{tabular}
\end{table}

\Tab~\ref{tab:cutflow} shows the numbers of events remaining after several stages of 
the selection.
A total of 117, 9 and 7293 candidate events were observed to pass all
but the \MET requirement of SR~A, B and C, respectively. After
imposing the final \MET requirement, no events remained for SR~A and B, while
two events remained for SR~C.

\begin{table}[htp]
  \caption{Samples of selected events at progressive stages of the selection. Where no number is shown
   the cut was not applied.}
  \label{tab:cutflow}
  \centering
  \small
  \begin{tabular}{|l||c|c|c| }
    \hline
    Triggered events  & \multicolumn{3}{|c|}{1166060} \\
    Diphoton selection     & \multicolumn{3}{|c|}{10455}  \\
    \hline
                 &  A & B & C \\
    \hline
    \dphim requirement & 7293 & -- & 7293 \\
    \HT  requirement        &  117 &  9 &   --   \\
    \MET  requirement       &    0 &  0 &    2  \\
    \hline
  \end{tabular}
\end{table}

\begin{figure}[tb]
\begin{center}
\includegraphics[width=0.48\textwidth]{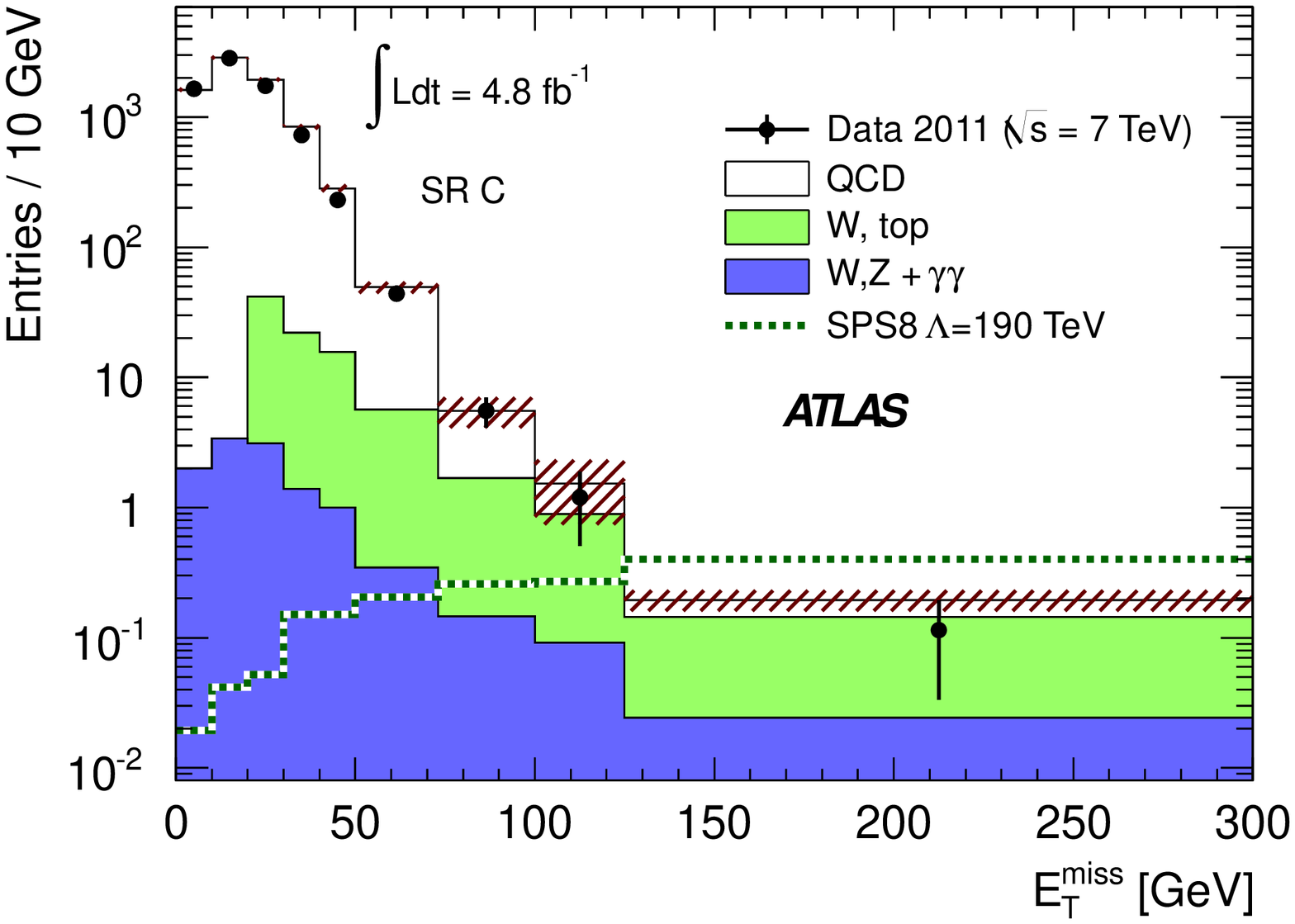}
\end{center}
\caption{\MET spectra in SR~C for the $\gamma\gamma$ candidate events
  in data (points, statistical uncertainty only) and the estimated QCD
  background (normalised to the number of $\gamma\gamma$ candidates
  with $\MET < \unit[20]{GeV}$), the $\myW$ and $\myttbar$ backgrounds
  as estimated from the electron--photon control sample, and the
  irreducible background of \Zgg and \Wgg. The hatched region represents
  the extent of the uncertainty on the
  total background prediction. Also shown is
  the expected signal from the SPS8 ($\Lambda = \unit[190]{TeV}$) sample.
  \label{fig:MET_C}
}
\end{figure}

Figure~\ref{fig:MET_C} shows the \MET distribution for SR C, the
expected contributions from the SPS8 MC sample with $\Lambda =
\unit[190]{TeV}$, and estimated background contributions from various
sources (described below). 

\section{Background estimation}
Following the procedure described in \Ref~\cite{Aad:2011kz}, the
contribution to the large~\MET diphoton sample from SM sources can be
grouped into three primary components.  The first of these, referred
to as ``QCD background'', arises from a mixture of
processes that include $\gamma \gamma$ production as well as $\gamma $
+ jet and multijet events with at least one jet mis-reconstructed as a
photon.  The second background component, referred to as ``EW background'', is due to $W+X$ and \ttbar
events (here ``$X$'' can be any number of photons or jets), and
where mis-reconstructed photons can arise from electrons and jets, and for
which final-state neutrinos produce significant \met. 
The QCD and EW backgrounds
were estimated via dedicated control samples of data events.
The third background component,
referred to as ``irreducible'',
consists of $W$ and $Z$ bosons produced in association with two 
real photons, with a subsequent decay into one or more neutrinos.

To estimate the QCD background from $\gamma\gamma$, $\gamma$
+ jet, and multijet events, a ``\QCDg control sample'' was selected
from the diphoton trigger sample by selecting events for which at
least one of the photon candidates passes the loose but not the tight photon
identification. Events with electrons were vetoed to remove contamination from
$W\to e\nu$ decays. 
The \HT and \dphi requirements associated with each of the three SRs were then applied,
yielding three separate \QCDg samples, or ``templates''.
An estimate of the QCD background contamination in each SR 
was obtained from imposing the \MET
requirement associated with the given SR upon the corresponding QCD template, after normalising each template 
to the diphoton data with $\MET < \unit[20]{GeV}$ from the given SR.
This yielded a QCD
background expectation of
$0.85\pm0.30(\mathrm{stat})$ events for SR C. No events above the
corresponding \MET requirement were observed for the A and B control samples,
yielding an estimate of 0 events with a 90\% confidence-level (CL) upper limit of
less than 1.01 and 1.15 background events for SR A and SR B, respectively.

To improve the constraint on the estimated background for SRs A and B,
a complementary method making use of \HT sidebands of the \QCDg control sample
was employed. The \HT requirement applied to the \QCDg templates of SR A and B
was relaxed in three steps: to \unit[400]{GeV}, \unit[200]{GeV} and \unit[0]{GeV} 
for the SR A control sample, and to \unit[800]{GeV}, \unit[400]{GeV}
and \unit[200]{GeV} for the SR B control sample. For each SR, the \MET distribution of each 
of these relaxed control samples was scaled to the diphoton \MET distribution
for $\MET < \unit[20]{GeV}$ of the given SR, yielding a series of three expected values for the QCD
background as a function of the applied \HT requirement. 
The complementary estimate for the 
background contribution to the signal region
employed a parabolic extrapolation to the actual \HT
requirement used for the analysis (600 GeV and 1100 GeV for SRs A and B, respectively); 
a linear fit yielded a significantly lower background estimate for both SRs. 
The parabolic extrapolation yielded conservative
upper estimates of 0.14 and 0.54 events for SRs A and B, respectively.
The overall QCD background estimates for SRs A and B were taken to be $0.07 \pm 0.07(\mathrm{syst})$ and
$0.27 \pm 0.27(\mathrm{syst})$ events, respectively, half of the value of this upper estimate, with
systematic uncertainty assigned to cover the entire range between 0 and the upper estimate.
The choice of a parabolic function constrained by three \HT points does not permit an estimation of 
statistical uncertainty on the extrapolation.

Other sources of systematic uncertainty in the estimated QCD background were considered.
Using the \MET distribution from a sample of $Z \rightarrow e^+ e^-$ events instead
of that of the \QCDg sample yielded estimates of 0, 0 and 0.15 events for the SRs
A, B and C, respectively. The difference between this estimate and
that of the \QCDg sample was incorporated as a systematic
uncertainty of $\pm 0.71$ on the SR C QCD background estimate.
Making use of the alternative ranges $\unit[5]{GeV} < \MET < \unit[25]{GeV}$ and $\unit[10]{GeV} < \MET < \unit[30]{GeV}$
over which the \QCDg sample was normalized to the 
$\gamma\gamma$ sample resulted in a further systematic uncertainty of $\pm 0.03$
events on the QCD background estimate for SR C. The resulting QCD background estimates
for the three SRs, along with their uncertainties, are compiled
in \Tab~\ref{tab:background_results}.

The EW background,
from $W+X$ and \ttbar
events, was estimated via an ``electron--photon'' control sample
composed of events with at least one tight photon and one electron, each with
$\ET>\unit[50]{GeV}$, and scaled by the probability for an electron to
be mis-reconstructed as a tight photon, as estimated from a ``tag-and-probe'' study of the
$Z$ boson in the $ee$ and $e\gamma$ sample. The scaling factor varies
between \unit[2.5]{\%} ($0 < |\eta| < 0.6$) and \unit[7.0]{\%} ($1.52 < |\eta| < 1.81$), 
since it depends on the amount of material in front of the calorimeter.
Events with two or more tight photons were vetoed from the control sample to preserve
its orthogonality to the signal sample. In case of more than one electron,
the one with the highest \pt was used.

After applying corresponding selection requirements on \HT, \dphi and \MET, a 
total of 1, 3 and 26 electron--photon events were observed for SRs 
A, B and C, respectively. After multiplying by the 
$\eta$-dependent electron-to-photon mis-reconstruction probability, the
resulting EW background contamination was estimated to be 
$0.03 \pm 0.03$, $0.09 \pm 0.05$ and $0.80 \pm 0.16$ events for
SRs A, B and C, respectively, where the 
uncertainties are statistical only.

\begin{table*}[t!]
  \caption{The expected number of $\gamma\gamma$ events for
    each of the three signal regions.
    The uncertainties are statistical,
    arising from the limited numbers of events in the control samples, and
    systematic, the details of which are given in the text. For the
    irreducible background, the statistical uncertainty is due to
    limited numbers of events in the corresponding MC samples.}
  \label{tab:background_results}
  \center
  \begin{tabular}{l|r|r|r}\hline
    & SR A  & SR B & SR C \\ \hline
      QCD                         & $ 0.07 \pm 0.00 \pm 0.07 $ & $ 0.27 \pm 0.00 \pm 0.27 $ &  $ 0.85 \pm 0.30 \pm 0.71 $  \\
      Electroweak                 &  $ 0.03 \pm 0.03 \pm 0.01 $ & $ 0.09 \pm 0.05 \pm 0.02 $ & $ 0.80  \pm 0.16 \pm 0.22 $ \\
      \Wgg  & $< 0.01$ & $< 0.01$ & $ 0.18  \pm 0.13 \pm 0.18 $ \\
      \Zgg & $< 0.01$ & $< 0.01$ & $0.27 \pm 0.09 \pm 0.04$  \\
      \hline
      Total & $0.10 \pm 0.03 \pm 0.07$ & $0.36 \pm 0.05 \pm 0.27$ & $2.11 \pm 0.37 \pm 0.77$ \\
      \hline
      Observed events  & 0 & 0 & 2 \\
      \hline
  \end{tabular}
\end{table*}

The systematic uncertainty on the determination of the electron-to-photon
mis-reconstruction probability is assessed by performing an
independent tag-and-probe analysis with looser electron \ET
and identification requirements. Differences with the
nominal tag-and-probe analysis are taken as systematic
uncertainty on the EW background estimate, resulting in relative systematic
uncertainties of $\pm 6.9$\%, $\pm 7.1$\% and $\pm 10.0$\% for SRs
A, B and C, respectively.
MC studies suggest that approximately 25\% of the EW background
involves no electron-to-photon mis-reconstruction, and
thus are not accounted for with the electron--photon
control sample. These events, however, typically involve
a jet-to-photon mis-reconstruction (for example, an event
with one radiated photon and a hadronic $\tau$ decay
with an energetic leading $\pi^0$ mis-reconstructed as a photon), 
and are thus potentially accounted for in the QCD background estimate.
A relative systematic uncertainty of $\pm 25$\% is conservatively assigned to 
the EW background estimates for all three SRs to account
for this ambiguity.
The resulting EW background estimates
for the three SRs, along with their uncertainties, are compiled
in \Tab~\ref{tab:background_results}.

The contribution of the irreducible background from the \Zgg and \Wgg
processes was estimated using MC samples. It was found to be
negligible for SRs A and B, and estimated to be $0.46 \pm 0.16 \pm
0.19$ events for SR C, where the first uncertainty is due to the
limited number of events in the MC sample and the second to the uncertainty
on the applied $K$-factor. These estimates, along with the resulting
estimates for the total background from all sources, are reported
in \Tab~\ref{tab:background_results}.

The contamination from cosmic-ray muons, estimated using events triggered
in empty LHC bunches, was found to be negligible.

\section{Signal efficiencies and systematic uncertainties}

Signal efficiencies were estimated using MC simulation.
GGM signal efficiencies were estimated over
an area of the GGM parameter space that ranges from \unit[800]{GeV} to
\unit[1300]{GeV} for the gluino or squark mass, and from
$\unit[50]{GeV}$ to within $\unit[10]{GeV}$ of the gluino or squark
mass for the neutralino mass. For SR A the efficiency increases
smoothly from \unit[1.2]{\%} to \unit[25]{\%} for
$(\mass{\gluino},\mass{\neutralino}) = \unit[(800,50)]{GeV}$ to
$\unit[(1300,1280)]{GeV}$,
but then drops to \unit[20]{\%} for
the case for which the gluino and neutralino masses are only separated by
\unit[10]{GeV}. 
For SR B the efficiency increases smoothly from \unit[2.8]{\%} to
\unit[26]{\%} for $(\mass{\gluino},\mass{\neutralino}) =
\unit[(800,790)]{GeV}$ to $\unit[(1300,50)]{GeV}$.
The SPS8 signal efficiency in SR~C increases smoothly from
\unit[5.9]{\%} ($\Lambda = \unit[100]{TeV}$) to \unit[21]{\%}
($\Lambda = \unit[250]{TeV}$). 
For SR~A the UED signal efficiency increases smoothly from \unit[28]{\%} ($1/R =
\unit[1.0]{TeV}$) to \unit[37]{\%} ($1/R = \unit[1.5]{TeV}$).

\begin{table}[b!]
  \caption{Relative systematic uncertainties on the expected signal
    yield for the GGM model with 
    $\mass{\gluino} = \unit[1000]{GeV} and \mass{\neutralino} = \unit[450]{GeV}$,
   the SPS8 model with $\Lambda=\unit[190]{TeV}$, and
    the UED model with $1/R = \unit[1.3]{TeV}$. For the GGM model, when the uncertainty 
differs for SRs A and B, it is presented as SRA/SRB.
No PDF and scale uncertainties
    are given for the UED case as the cross section is evaluated only to
    LO. }
  \label{tab:systematics}
  \center
  \begin{tabular}{lrrr} 
    \hline
    Source of uncertainty & \multicolumn{3}{c}{Uncertainty} \\
    & \multicolumn{1}{c}{GGM} & \multicolumn{1}{c}{SPS8} & \multicolumn{1}{c}{UED} \\ \hline
    Integrated luminosity           & $\hspace{0.04in}3.9$\% & $\hspace{0.04in}3.9$\% & $\hspace{0.04in}3.9$\% \\
    Trigger                         & $\hspace{0.04in}0.5$\% & $\hspace{0.04in}0.5$\% & $\hspace{0.04in}0.5$\% \\
    Photon identification           & $\hspace{0.02in}4.4$\% & $\hspace{0.02in}4.4$\% & $\hspace{0.02in}4.4$\% \\
    Photon isolation                & $\hspace{0.02in}0.9$\% & $\hspace{0.02in}0.2$\% & $\hspace{0.02in}0.4$\% \\
    Pile-up                         & $\hspace{0.02in}0.8$\% & $\hspace{0.02in}0.5$\% & $\hspace{0.02in}0.5$\% \\
    $\MET$ reconstruction           & $\hspace{0.02in}3.9/1.1$\% & $\hspace{0.02in}2.8$\% & $\hspace{0.02in}1.5$\% \\
    $\HT$                           & $0.0/2.1$\% & $-$ & $0.4$\% \\
    Signal MC sample size           & $\hspace{0.02in}3.0$\% & $\hspace{0.02in}2.1$\% & $\hspace{0.02in}1.4$\% \\
    \hline
    Total signal uncertainty        & $\hspace{0.02in}7.6/7.1$\% & $\hspace{0.02in}6.8$\%&  $\hspace{0.02in}6.3$\% \\
    \hline
    PDF and scale                   & $\hspace{0.02in}31$\% & $\hspace{0.02in}5.5$\% & $\hspace{0.02in}-$  \\ 
    \hline
    Total                           & $\hspace{0.02in}32$\% & $\hspace{0.02in}8.7$\% & $\hspace{0.02in}6.3$\% \\
    \hline
  \end{tabular}
\end{table}

The various relative systematic uncertainties on the GGM, SPS8 and UED
signal cross sections are summarised in \Tab~\ref{tab:systematics} for
the chosen reference points: $(\mass{\gluino},\mass{\neutralino}) =
    \unit[(1000,450)]{GeV}$ for GGM, $\Lambda=\unit[190]{TeV}$ for SPS8, and
    $1/R = \unit[1.3]{TeV}$ for UED. The uncertainty on the
luminosity is $\pm \unit[3.9]{\%}$~\cite{lumi2011,Aad:2011dr}.
The efficiency of the required diphoton trigger was estimated using 
a single photon trigger, the efficiency of which was determined
using a bootstrap method~\cite{Aad:2012xs}. The result
is $99.8^{+0.2}_{-0.8} \%$ for events passing the diphoton selection.
To estimate the systematic uncertainty due to the unknown composition
of the data sample, the trigger efficiency was also evaluated on MC
events using mis-reconstructed photons from filtered multijet samples
and photons from signal (GGM, SPS8 and UED) samples. A conservative
systematic uncertainty of $\pm \unit[0.5]{\%}$ was derived from the
difference between the obtained efficiencies. Uncertainties on the
photon selection, the photon energy scale, and the detailed material
composition of the detector, as described in \Ref~\cite{Aad:2011kz},
result in an uncertainty of $\pm \unit[4.4]{\%}$ for the GGM, SPS8 and UED
signals. The uncertainty due to the photon isolation requirement was estimated by
varying the energy leakage and the pile-up corrections independently,
resulting in an uncertainty of $\pm \unit[0.9]{\%}$, $\pm \unit[0.2]{\%}$ and
$\pm \unit[0.4]{\%}$ for the GGM, SPS8 and UED signals, respectively. The influence of
pile-up on the signal efficiency, evaluated by scaling the number of
pile-up events in the MC simulation by a factor of 0.9 (chosen to
reflect the range of uncertainty inherent in estimating and modeling
the effects of pile-up) leads to a systematic
uncertainty of $\pm \unit[0.8]{\%}$~(GGM), $\pm \unit[0.5]{\%}$~(SPS8) and
$\pm \unit[0.5]{\%}$~(UED). Systematic uncertainties due to the $\MET$
reconstruction, estimated by varying the cluster energies and the
$\MET$ resolution between the measured performance and MC
expectations~\cite{Aad:2012re}, 
contribute an uncertainty of $\pm \unit[0.1/0.5]{\%}$ to
$\pm \unit[5.3/16.1]{\%}$~(GGM, SR A/B), $\pm \unit[1.6]{\%}$ to $\pm \unit[9.7]{\%}$
(SPS8) and $\pm \unit[0.9]{\%}$ to $\pm \unit[2]{\%}$~(UED). Systematic
uncertainties due to the \HT reconstruction, estimated by varying the
energy scale and resolution of the individual objects entering \HT,
are below $\pm \unit[0.3]{\%}$~(GGM, SR A), $\pm \unit[0.1]{\%}$ to
$\pm \unit[7.3]{\%}$~(GGM, SR B) and $\pm \unit[1.1]{\%}$ to
$\pm \unit[0.1]{\%}$~(UED). The systematic uncertainties from \MET and \HT
are taken to be fully correlated.
Added in quadrature, the total systematic uncertainty on the signal
yield varies between $\pm \unit[6]{\%}$ and $\pm \unit[20]{\%}$~(GGM),
$\pm \unit[6]{\%}$ and $\pm \unit[15]{\%}$~(SPS8), and $\pm \unit[6]{\%}$ and
$\pm \unit[7]{\%}$ (UED).

The PDF and factorisation and renormalisation scale uncertainties 
on the GGM (SPS8) cross sections were evaluated as described in
Section 4, leading to a
combined systematic uncertainty between $\pm \unit[23$--$39]{\%}$, 
$\pm \unit[29$--$49]{\%}$ and $\pm \unit[4.7$--$6.4]{\%}$ for the GGM (gluino), 
GGM (squark) and SPS8 models, respectively. 
The different impact of the PDF and scale uncertainties on the GGM and
SPS8 yields is related to the different production mechanisms in the
two models (see \Section~\ref{sec:susy}).
In the case of UED, the PDF uncertainties were evaluated by using the
{\tt MSTW2008 LO}~\cite{MSTW2008} PDF error sets in the LO 
cross-section calculation and are about $\pm \unit[4]{\%}$. The scale of
$\alpha_s$ in the LO cross section calculation was increased and
decreased by a factor of two, leading to a systematic uncertainty of
$\pm \unit[4.5]{\%}$ and $\pm \unit[9]{\%}$, respectively. 
NLO calculations are not yet available, 
so the LO cross sections were used for the
limit calculation without any theoretical uncertainty, and the effect
of PDF and scale uncertainties on the final limit is given separately.

\begin{figure}[t]
  \begin{center}
    \includegraphics[width=0.55\textwidth]{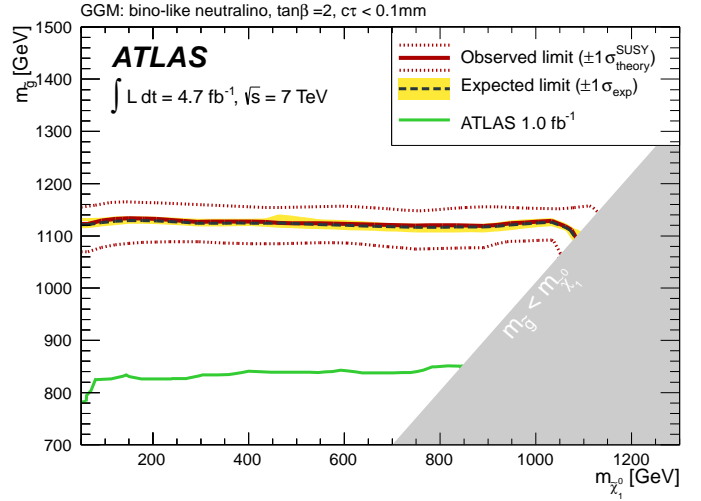}
  \end{center}
  \caption{Expected and observed \unit[95]{\%} \CL lower limits on the
    gluino mass as a function of the neutralino mass in the GGM model
    with a bino-like lightest neutralino NLSP (the grey area indicates
    the region for which the gluino mass is less than the bino mass, which is not considered
    here). The other sparticle masses are 
    assumed to be decoupled.
    Further model parameters are $\tan\beta=2$
    and $c\tau_{\mathrm{NLSP}} < \unit[0.1]{mm}$. The previous
    ATLAS limit~\protect\cite{Aad:2011zj} is also shown.
    \label{fig:ggm:limit:g}
  }
\end{figure}

\begin{figure}[t]
  \begin{center}
    \includegraphics[width=0.55\textwidth]{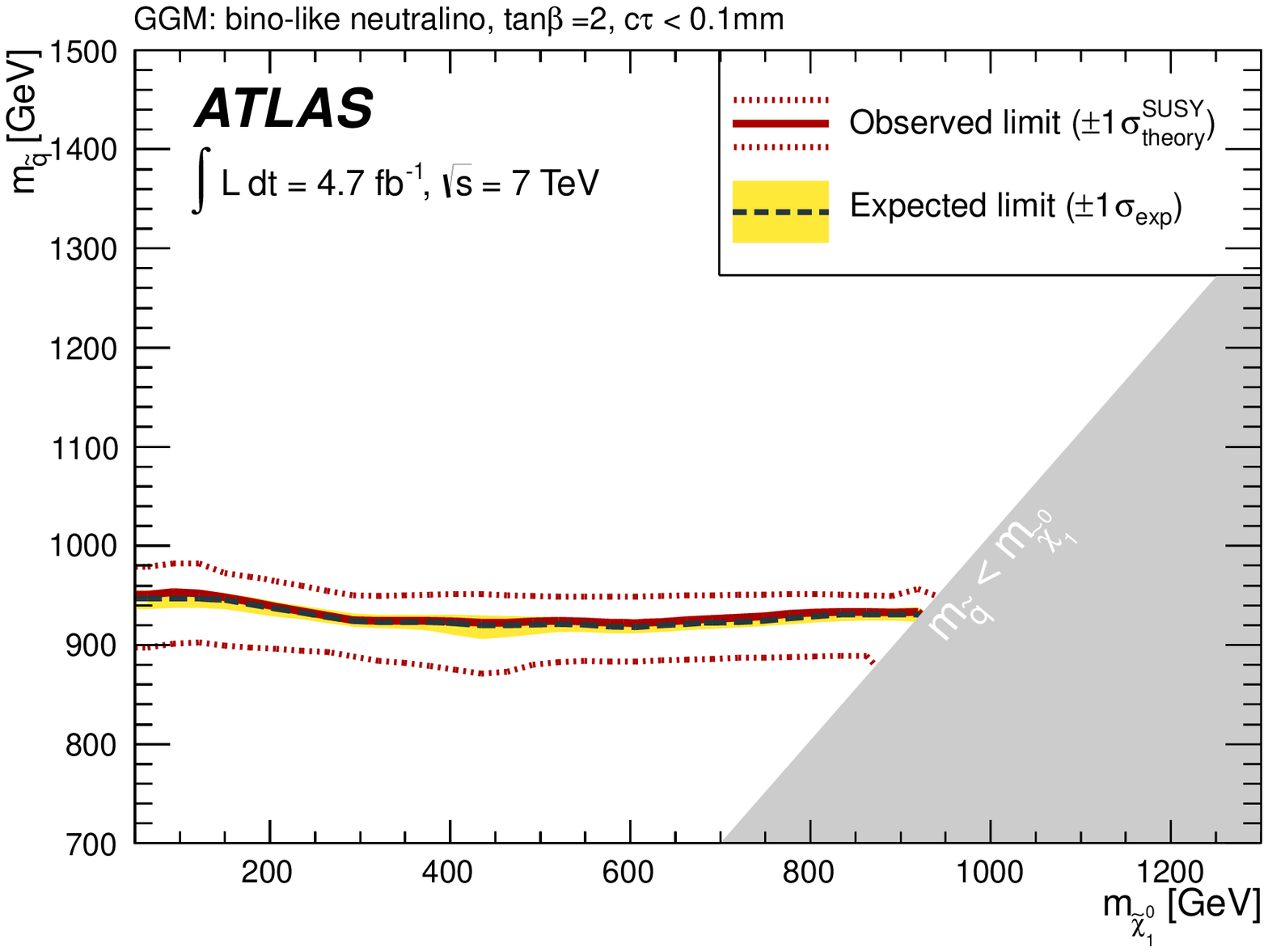}
  \end{center}
  \caption{Expected and observed \unit[95]{\%} \CL lower limits on the
    squark mass as a function of the neutralino mass in the GGM model
    with a bino-like lightest neutralino NLSP 
    (the grey area indicates the region for which the squark mass is less than 
     the bino mass, which is not considered here).
    The other sparticle masses are 
    assumed to be decoupled.
    Further model parameters are $\tan\beta=2$
    and $c\tau_{\mathrm{NLSP}} < \unit[0.1]{mm}$.
    \label{fig:ggm:limit:sq}
  }
\end{figure}

\begin{figure}[t]
  \begin{center}
    \includegraphics[width=0.44\textwidth]{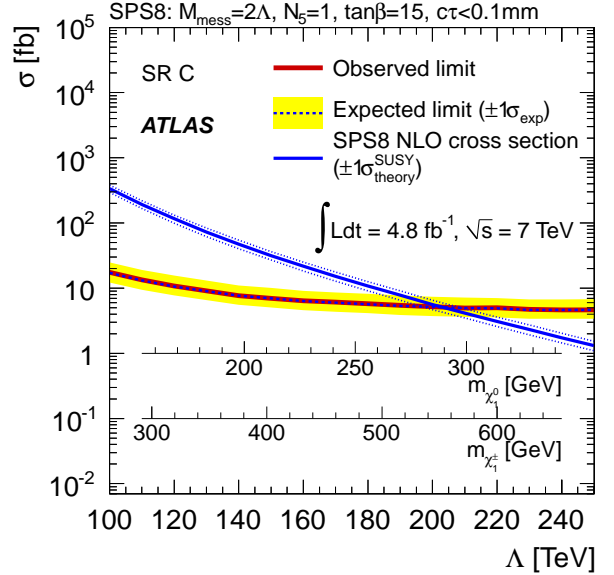}
  \end{center}
  \caption{Expected and observed \unit[95]{\%}~\CL upper limits on the
    sparticle production cross section in the SPS8 model, and the NLO
    cross-section prediction, as a function of $\Lambda$ and the
    lightest neutralino and chargino masses.  Further SPS8 model
    parameters are $M_\mathrm{mess}=2\Lambda$, $\Nfive=1$,
    $\tan\beta=15$ and $c\tau_{\mathrm{NLSP}} < \unit[0.1]{mm}$.
    Limits are set based on SR C.
      \label{fig:sps8:limit}
    }
\end{figure}

\begin{figure}[t]
  \begin{center}
    \includegraphics[width=0.44\textwidth]{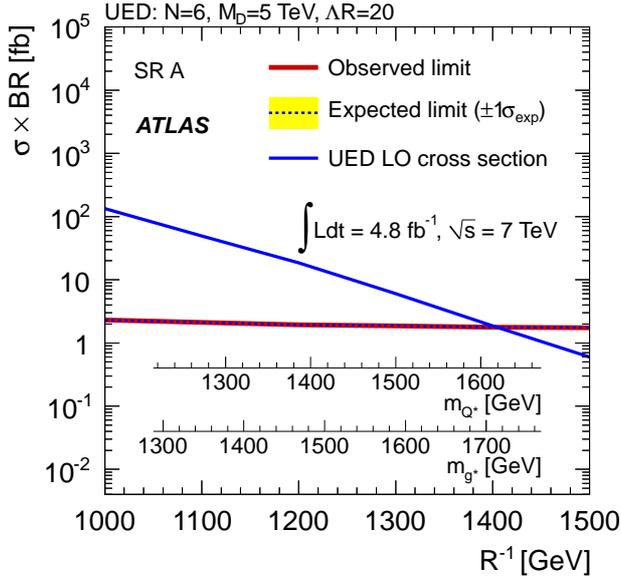}
  \end{center}
  \caption{Expected and observed \unit[95]{\%}~\CL upper limits on the
    KK particle production cross section times branching ratio to
    two photons in the UED model, and the LO cross-section prediction
    times branching ratio, as a function of $1/R$ and the KK quark
    ($Q^{*}$) and KK gluon ($g^{*}$) masses. 
    The $\pm 1 \sigma$
    expected-limit error band 
    overlaps the observed limit contour and is too
    narrow to be distinguished.
    No error is shown for the UED cross section since the cross-section calculation
    is available only to LO (see text for further discussion).
    The UED model parameters
    are $N=6$, $M_D = \unit[5]{TeV}$ and $\Lambda R = 20$. Limits are
    set based on SR A.}
  \label{fig:ued:limit}
\end{figure}

\section{Results}
No evidence for physics beyond the SM was observed in any of the SRs.
Based on the numbers of observed events in SR~A, B and C and the
background expectation shown in \Tab~\ref{tab:background_results},
\unit[95]{\%}~\CL upper limits are set on the numbers of events in the
different SRs from any scenario of physics beyond the SM using the
profile likelihood and $CL_s$ prescriptions~\cite{Read:2002hq}.
Uncertainties
on the background and signal expectations are treated as
Gaussian-distributed nuisance parameters in the maximum likelihood fit, 
resulting in observed upper limits of 3.1, 3.1 and 4.9 events
for SRs A, B and C, respectively. 
These limits translate into \unit[95]{\%} upper
limits on the visible cross section for new physics, defined by the
product of cross section, branching ratio, acceptance and
efficiency for the different SR definitions, 
of 0.6, 0.6 and 1.0 fb, respectively. Because the observed
numbers of events are close to the expected numbers of
background events for all three SRs, expected limits on the 
numbers of events from and visible cross section for new physics
are, to the quoted accuracy, identical to the observed limits.

Limits are also set on the GGM squark and gluino masses as a function
of the bino-like neutralino mass, making use of the SR (A or B) that provides
the most stringent expected limit for the given neutralino mass.
Figures~\ref{fig:ggm:limit:g} and \ref{fig:ggm:limit:sq} show the
expected and observed lower limits on the GGM gluino and squark masses, respectively, as a
function of the neutralino mass. Three observed-limit contours are shown, 
corresponding to the nominal assumption for the SUSY production cross section
as well as those derived by reducing and increasing the cross 
section by one standard deviation of theoretical uncertainty (the combined
uncertainty due to the PDFs and renormalisation and factorisation scales). 
For comparison the lower limits on the GGM gluino mass from
ATLAS~\cite{Aad:2011zj} based on \unit[1]{\ifb} from 2011 are also
shown. 

Including all sources of uncertainty other than 
the theoretical uncertainty,
\unit[95]{\%}~\CL upper limits on the cross section
of the SPS8 model are derived 
from the SR C result
and displayed in
\Fig~\ref{fig:sps8:limit} for the range $\Lambda=\unit[100$--$250]{TeV}$
along with the overall production cross section and its theoretical uncertainty.
For illustration the cross-section
dependence as a function of the lightest neutralino and chargino
masses is also shown. 

Figure~\ref{fig:ued:limit} shows the
limit on the cross section times branching ratio for the UED model 
as a function of the compactification scale $1/R$, derived from the result of SR A.
A \unit[95]{\%}~\CL lower limit of $1/R > \UEDlimitlow$ is set.
For illustration the cross-section dependence as
a function of the KK quark and KK gluon masses is also shown.
Again, neither PDF nor scale uncertainties are included when
calculating the limits; including PDF and scale uncertainties, computed
at LO, in the limit calculation degrades the limit on $1/R$ by a few \unit{GeV}.

\section{Conclusions}
A search for events with two photons and substantial \MET,
performed using \integLumi of \unit[7]{TeV} $pp$ collision data
recorded with the ATLAS detector at the LHC, is presented. The
sensitivity to different new physics models producing this final
state was optimised by defining three different SRs. No significant
excess above the expected background is found in any SR.
The results are used to set model-independent \unit[95]{\%}~\CL
upper limits on possible contributions from new physics. 
In addition, under the GGM hypothesis, considering cross sections one standard
deviation of theoretical uncertainty below the nominal value,
a lower limit on the gluino/squark mass of
\GGMlimitg/\GGMlimitsq is determined for bino masses above
\unit[50]{GeV}. Under similar assumptions, a lower limit of \SPSlimit is set on the SUSY-breaking scale
scale $\Lambda$ of the SPS8 model. Considering nominal values of the leading-order UED cross section, 
a lower limit of \UEDlimitlow is set on the UED
compactification scale $1/R$. These results provide the most stringent
tests of these models to date.

\input{Acknowledgement-31Aug12-collisions.tex}

\bibliographystyle{atlasnote}
\bibliography{diphoton}

\onecolumn
\clearpage
\input{atlas_authlist}

\end{document}

%% file: Acknowledgement-31Aug12-collisions.tex



\section{Acknowledgements}

We thank CERN for the very successful operation of the LHC, as well as the
support staff from our institutions without whom ATLAS could not be
operated efficiently.

We acknowledge the support of ANPCyT, Argentina; YerPhI, Armenia; ARC,
Australia; BMWF and FWF, Austria; ANAS, Azerbaijan; SSTC, Belarus; CNPq and FAPESP,
Brazil; NSERC, NRC and CFI, Canada; CERN; CONICYT, Chile; CAS, MOST and NSFC,
China; COLCIENCIAS, Colombia; MSMT CR, MPO CR and VSC CR, Czech Republic;
DNRF, DNSRC and Lundbeck Foundation, Denmark; EPLANET and ERC, European Union;
IN2P3-CNRS, CEA-DSM/IRFU, France; GNSF, Georgia; BMBF, DFG, HGF, MPG and AvH
Foundation, Germany; GSRT, Greece; ISF, MINERVA, GIF, DIP and Benoziyo Center,
Israel; INFN, Italy; MEXT and JSPS, Japan; CNRST, Morocco; FOM and NWO,
Netherlands; RCN, Norway; MNiSW, Poland; GRICES and FCT, Portugal; MERYS
(MECTS), Romania; MES of Russia and ROSATOM, Russian Federation; JINR; MSTD,
Serbia; MSSR, Slovakia; ARRS and MVZT, Slovenia; DST/NRF, South Africa;
MICINN, Spain; SRC and Wallenberg Foundation, Sweden; SER, SNSF and Cantons of
Bern and Geneva, Switzerland; NSC, Taiwan; TAEK, Turkey; STFC, the Royal
Society and Leverhulme Trust, United Kingdom; DOE and NSF, United States of
America.

The crucial computing support from all WLCG partners is acknowledged
gratefully, in particular from CERN and the ATLAS Tier-1 facilities at
TRIUMF (Canada), NDGF (Denmark, Norway, Sweden), CC-IN2P3 (France),
KIT/GridKA (Germany), INFN-CNAF (Italy), NL-T1 (Netherlands), PIC (Spain),
ASGC (Taiwan), RAL (UK) and BNL (USA) and in the Tier-2 facilities
worldwide.


%% file: atlas_authlist.tex
\begin{flushleft}
{\Large The ATLAS Collaboration}

\bigskip

G.~Aad$^{\rm 48}$,
T.~Abajyan$^{\rm 21}$,
B.~Abbott$^{\rm 111}$,
J.~Abdallah$^{\rm 12}$,
S.~Abdel~Khalek$^{\rm 115}$,
A.A.~Abdelalim$^{\rm 49}$,
O.~Abdinov$^{\rm 11}$,
R.~Aben$^{\rm 105}$,
B.~Abi$^{\rm 112}$,
M.~Abolins$^{\rm 88}$
% ...
\bigskip

% ...
$^{11}$ Institute of Physics, Azerbaijan Academy of Sciences, Baku, Azerbaijan\\
$^{12}$ Institut de F\'{i}sica d'Altes Energies and Departament de F\'{i}sica de la Universitat Aut\`{o}noma de Barcelona and ICREA, Barcelona, Spain\\
% ...
$^{21}$ Physikalisches Institut, University of Bonn, Bonn, Germany\\
% ...
$^{48}$ Fakult\"{a}t f\"{u}r Mathematik und Physik, Albert-Ludwigs-Universit\"{a}t, Freiburg, Germany\\
$^{49}$ Section de Physique, Universit\'{e} de Gen\`{e}ve, Geneva, Switzerland\\
% ...
$^{88}$ Department of Physics and Astronomy, Michigan State University, East Lansing MI, United States of America\\
% ...
$^{105}$ Nikhef National Institute for Subatomic Physics and University of Amsterdam, Amsterdam, Netherlands\\
% ...
$^{111}$ Homer L. Dodge Department of Physics and Astronomy, University of Oklahoma, Norman OK, United States of America\\
$^{112}$ Department of Physics, Oklahoma State University, Stillwater OK, United States of America\\
% ...
$^{115}$ LAL, Universit\'{e} Paris-Sud and CNRS/IN2P3, Orsay, France\\
% ...
\end{flushleft}


%% file: diphoton.bbl
\providecommand{\href}[2]{#2}\begingroup\raggedright\begin{thebibliography}{10}

\bibitem{Aad:2011zj}
{ATLAS} Collaboration,  Phys. Lett. {\bf B710} (2012)  519,
\href{http://arxiv.org/abs/1111.4116}{{\tt arXiv:1111.4116 [hep-ex]}}.

\bibitem{Meade:2008wd}
P.~Meade, N.~Seiberg, and D.~Shih,
  \href{http://dx.doi.org/10.1143/PTPS.177.143}{Prog. Theor. Phys. Suppl. {\bf
  177} (2009)  143},
\href{http://arxiv.org/abs/0801.3278}{{\tt arXiv:0801.3278 [hep-ph]}}.

\bibitem{Buican:2008ws}
M.~Buican, P.~Meade, N.~Seiberg, and D.~Shih,
  \href{http://dx.doi.org/10.1088/1126-6708/2009/03/016}{JHEP {\bf 03} (2009)
  016},
\href{http://arxiv.org/abs/0812.3668}{{\tt arXiv:0812.3668 [hep-ph]}}.

\bibitem{Ruderman:2011vv}
J.~T. Ruderman and D.~Shih,
  \href{http://dx.doi.org/10.1007/JHEP08(2012)159}{JHEP {\bf 1208} (2012)
  159},
\href{http://arxiv.org/abs/1103.6083}{{\tt arXiv:1103.6083 [hep-ph]}}.

\bibitem{Allanach:2002nj}
B.~C. Allanach et al.,  \href{http://dx.doi.org/10.1007/s10052-002-0949-3}{Eur.
  Phys. J. {\bf C25} (2002)  113},
\href{http://arxiv.org/abs/hep-ph/0202233}{{\tt arXiv:hep-ph/0202233}}.

\bibitem{Appelquist:2000nn}
T.~Appelquist, H.-C. Cheng, and B.~A. Dobrescu,
  \href{http://dx.doi.org/10.1103/PhysRevD.64.035002}{Phys. Rev. {\bf D64}
  (2001)  035002},
\href{http://arxiv.org/abs/hep-ph/0012100}{{\tt arXiv:hep-ph/0012100}}.

\bibitem{Macesanu:2002ew}
C.~Macesanu, C.~McMullen, and S.~Nandi,
  \href{http://dx.doi.org/10.1016/S0370-2693(02)02694-1}{Phys. Lett. {\bf B546}
  (2002)  253}, \href{http://arxiv.org/abs/hep-ph/0207269}{{\tt
  arXiv:hep-ph/0207269}}.

\bibitem{Macesanu:2005jx}
C.~Macesanu,  \href{http://dx.doi.org/10.1142/S0217751X06030886}{Int. J. Mod.
  Phys. {\bf A21} (2006)  2259},
  \href{http://arxiv.org/abs/hep-ph/0510418}{{\tt arXiv:hep-ph/0510418}}.

\bibitem{Miyazawa:1966}
H.~Miyazawa,
\href{http://dx.doi.org/10.1143/PTP.36.1266}{Prog. Theor. Phys. {\bf 36 (6)}
  (1966)  1266}.

\bibitem{Ramond:1971gb}
P.~Ramond,
\href{http://dx.doi.org/10.1103/PhysRevD.3.2415}{Phys. Rev. {\bf D3} (1971)
  2415}.

\bibitem{Golfand:1971iw}
Y.~A. Gol'fand and E.~P. Likhtman,  JETP Lett. {\bf 13} (1971)  323.
[Pisma Zh.Eksp.Teor.Fiz.13:452-455,1971].

\bibitem{Neveu:1971rx}
A.~Neveu and J.~H. Schwarz,
\href{http://dx.doi.org/10.1016/0550-3213(71)90448-2}{Nucl. Phys. {\bf B31}
  (1971)  86}.

\bibitem{Neveu:1971iv}
A.~Neveu and J.~H. Schwarz,
\href{http://dx.doi.org/10.1103/PhysRevD.4.1109}{Phys. Rev. {\bf D4} (1971)
  1109}.

\bibitem{Gervais:1971ji}
J.~Gervais and B.~Sakita,
\href{http://dx.doi.org/10.1016/0550-3213(71)90351-8}{Nucl. Phys. {\bf B34}
  (1971)  632}.

\bibitem{Volkov:1973ix}
D.~V. Volkov and V.~P. Akulov,
\href{http://dx.doi.org/10.1016/0370-2693(73)90490-5}{Phys. Lett. {\bf B46}
  (1973)  109}.

\bibitem{Wess:1973kz}
J.~Wess and B.~Zumino,
\href{http://dx.doi.org/10.1016/0370-2693(74)90578-4}{Phys. Lett. {\bf B49}
  (1974)  52}.

\bibitem{Wess:1974tw}
J.~Wess and B.~Zumino,
\href{http://dx.doi.org/10.1016/0550-3213(74)90355-1}{Nucl. Phys. {\bf B70}
  (1974)  39}.

\bibitem{Fayet:1976et}
P.~Fayet,
\href{http://dx.doi.org/10.1016/0370-2693(76)90319-1}{Phys. Lett. {\bf B64}
  (1976)  159}.

\bibitem{Fayet:1977yc}
P.~Fayet,
\href{http://dx.doi.org/10.1016/0370-2693(77)90852-8}{Phys. Lett. {\bf B69}
  (1977)  489}.

\bibitem{Farrar:1978xj}
G.~R. Farrar and P.~Fayet,
\href{http://dx.doi.org/10.1016/0370-2693(78)90858-4}{Phys. Lett. {\bf B76}
  (1978)  575}.

\bibitem{Fayet:1979sa}
P.~Fayet,
\href{http://dx.doi.org/10.1016/0370-2693(79)91229-2}{Phys. Lett. {\bf B84}
  (1979)  416}.

\bibitem{Dimopoulos:1981zb}
S.~Dimopoulos and H.~Georgi,
\href{http://dx.doi.org/10.1016/0550-3213(81)90522-8}{Nucl. Phys. {\bf B193}
  (1981)  150}.

\bibitem{Dine:1981gu}
M.~Dine and W.~Fischler,
\href{http://dx.doi.org/10.1016/0370-2693(82)91241-2}{Phys. Lett. {\bf B110}
  (1982)  227}.

\bibitem{AlvarezGaume:1981wy}
L.~Alvarez-Gaume, M.~Claudson, and M.~B. Wise,
\href{http://dx.doi.org/10.1016/0550-3213(82)90138-9}{Nucl. Phys. {\bf B207}
  (1982)  96}.

\bibitem{Nappi:1982hm}
C.~R. Nappi and B.~A. Ovrut,
\href{http://dx.doi.org/10.1016/0370-2693(82)90418-X}{Phys. Lett. {\bf B113}
  (1982)  175}.

\bibitem{Dine:1993yw}
M.~Dine and A.~E. Nelson,
  \href{http://dx.doi.org/10.1103/PhysRevD.48.1277}{Phys. Rev. {\bf D48} (1993)
   1277},
\href{http://arxiv.org/abs/hep-ph/9303230}{{\tt arXiv:hep-ph/9303230}}.

\bibitem{Dine:1994vc}
M.~Dine, A.~E. Nelson, and Y.~Shirman,
  \href{http://dx.doi.org/10.1103/PhysRevD.51.1362}{Phys. Rev. {\bf D51} (1995)
   1362},
\href{http://arxiv.org/abs/hep-ph/9408384}{{\tt arXiv:hep-ph/9408384}}.

\bibitem{Dine:1995ag}
M.~Dine, A.~E. Nelson, Y.~Nir, and Y.~Shirman,
  \href{http://dx.doi.org/10.1103/PhysRevD.53.2658}{Phys. Rev. {\bf D53} (1996)
   2658},
\href{http://arxiv.org/abs/hep-ph/9507378}{{\tt arXiv:hep-ph/9507378}}.

\bibitem{Aad:2012ag}
{ATLAS Collaboration},
  \href{http://dx.doi.org/10.1016/j.physletb.2012.06.055}{Phys.Lett. {\bf B714}
  (2012)  180},
\href{http://arxiv.org/abs/1203.6580}{{\tt arXiv:1203.6580 [hep-ex]}}.

\bibitem{Aad:2012rt}
{ATLAS Collaboration},  Phys. Lett. {\bf B714} (2012)  197,
\href{http://arxiv.org/abs/1204.3852}{{\tt arXiv:1204.3852 [hep-ex]}}.

\bibitem{Chatrchyan:2011ah}
{CMS} Collaboration,  \href{http://dx.doi.org/10.1007/JHEP06(2011)093}{JHEP
  {\bf 1106} (2011)  093}, \href{http://arxiv.org/abs/1105.3152}{{\tt
  arXiv:1105.3152 [hep-ex]}}.

\bibitem{Macesanu:2002db}
C.~Macesanu, C.~McMullen, and S.~Nandi,
  \href{http://dx.doi.org/10.1103/PhysRevD.66.015009}{Phys. Rev. {\bf D66}
  (2002)  015009}, \href{http://arxiv.org/abs/hep-ph/0201300}{{\tt
  arXiv:hep-ph/0201300}}.

\bibitem{DeRujula:2000he}
A.~De~Rujula, A.~Donini, M.~Gavela, and S.~Rigolin,
  \href{http://dx.doi.org/10.1016/S0370-2693(00)00479-2}{Phys. Lett. {\bf B482}
  (2000)  195}, \href{http://arxiv.org/abs/hep-ph/0001335}{{\tt
  arXiv:hep-ph/0001335}}.

\bibitem{Cheng:2002iz}
H.-C. Cheng, K.~T. Matchev, and M.~Schmaltz,
  \href{http://dx.doi.org/10.1103/PhysRevD.66.036005}{Phys. Rev. {\bf D66}
  (2002)  036005},
\href{http://arxiv.org/abs/hep-ph/0204342}{{\tt arXiv:hep-ph/0204342
  [hep-ph]}}.

\bibitem{ElKacimi:2009zj}
M.~ElKacimi, D.~Goujdami, H.~Przysiezniak, and P.~Z. Skands,
  \href{http://dx.doi.org/10.1016/j.cpc.2009.08.008}{Comput. Phys. Commun. {\bf
  181} (2010)  122}, \href{http://arxiv.org/abs/0901.4087}{{\tt arXiv:0901.4087
  [hep-ph]}}.

\bibitem{Djouadi:2002ze}
A.~Djouadi, J.-L. Kneur, and G.~Moultaka,
  \href{http://dx.doi.org/10.1016/j.cpc.2006.11.009}{Comput. Phys. Commun. {\bf
  176} (2007)  426}, \href{http://arxiv.org/abs/hep-ph/0211331}{{\tt
  arXiv:hep-ph/0211331}}.

\bibitem{Muhlleitner:2003vg}
M.~Muhlleitner, A.~Djouadi, and Y.~Mambrini,
  \href{http://dx.doi.org/10.1016/j.cpc.2005.01.012}{Comput. Phys. Commun. {\bf
  168} (2005)  46}, \href{http://arxiv.org/abs/hep-ph/0311167}{{\tt
  arXiv:hep-ph/0311167}}.

\bibitem{isajet}
F.~E. Paige, S.~D. Protopopescu, H.~Baer, and X.~Tata, {\em {ISAJET 7.69: A
  Monte Carlo Event Generator for $pp$, $\bar{p} p$, and $e^+e^-$
  Reactions}\/},
\href{http://arxiv.org/abs/hep-ph/0312045}{{\tt arXiv:hep-ph/0312045}}.

\bibitem{Bahr:2008pv}
M.~Bahr et al.,  \href{http://dx.doi.org/10.1140/epjc/s10052-008-0798-9}{Eur.
  Phys. J. {\bf C58} (2008)  639},
\href{http://arxiv.org/abs/0803.0883}{{\tt arXiv:0803.0883 [hep-ph]}}.

\bibitem{Sherstnev:2007nd}
A.~Sherstnev and R.~S. Thorne,
  \href{http://dx.doi.org/10.1140/epjc/s10052-008-0610-x}{Eur. Phys. J. {\bf
  C55} (2008)  553},
\href{http://arxiv.org/abs/0711.2473}{{\tt arXiv:0711.2473 [hep-ph]}}.

\bibitem{Beenakker:1996ch}
W.~Beenakker, R.~Hopker, M.~Spira, and P.~Zerwas,
  \href{http://dx.doi.org/10.1016/S0550-3213(97)00084-9}{Nucl. Phys. {\bf B492}
  (1997)  51}, \href{http://arxiv.org/abs/hep-ph/9610490}{{\tt
  arXiv:hep-ph/9610490}}.

\bibitem{Kulesza:2008jb}
A.~Kulesza and L.~Motyka,
  \href{http://dx.doi.org/10.1103/PhysRevLett.102.111802}{Phys. Rev. Lett. {\bf
  102} (2009)  111802},
\href{http://arxiv.org/abs/0807.2405}{{\tt arXiv:0807.2405 [hep-ph]}}.

\bibitem{Kulesza:2009kq}
A.~Kulesza and L.~Motyka,
  \href{http://dx.doi.org/10.1103/PhysRevD.80.095004}{Phys. Rev. {\bf D80}
  (2009)  095004},
\href{http://arxiv.org/abs/0905.4749}{{\tt arXiv:0905.4749 [hep-ph]}}.

\bibitem{Beenakker:2009ha}
W.~Beenakker, S.~Brensing, M.~Kramer, A.~Kulesza, E.~Laenen, et al.,
  \href{http://dx.doi.org/10.1088/1126-6708/2009/12/041}{JHEP {\bf 0912} (2009)
   041},
\href{http://arxiv.org/abs/0909.4418}{{\tt arXiv:0909.4418 [hep-ph]}}.

\bibitem{Beenakker:2011fu}
W.~Beenakker, S.~Brensing, M.~Kramer, A.~Kulesza, E.~Laenen, et al.,
  \href{http://dx.doi.org/10.1142/S0217751X11053560}{Int. J. Mod. Phys. {\bf
  A26} (2011)  2637--2664},
\href{http://arxiv.org/abs/1105.1110}{{\tt arXiv:1105.1110 [hep-ph]}}.

\bibitem{Kramer:2012bx}
M.~Kramer, A.~Kulesza, R.~van~der Leeuw, M.~Mangano, S.~Padhi, et al., {\em
  {Supersymmetry Production Cross Sections in pp Collisions at $\sqrt{s} = 7$
  TeV}\/},
\href{http://arxiv.org/abs/1206.2892}{{\tt arXiv:1206.2892 [hep-ph]}}.

\bibitem{pythia}
T.~Sjostrand, S.~Mrenna, and P.~Skands,  JHEP {\bf 05} (2006)  026,
\href{http://arxiv.org/abs/hep-ph/0603175}{{\tt arXiv:hep-ph/0603175}}.

\bibitem{Alwall:2007st}
J.~Alwall et al.,  \href{http://dx.doi.org/10.1088/1126-6708/2007/09/028}{JHEP
  {\bf 09} (2007)  028},
\href{http://arxiv.org/abs/0706.2334}{{\tt arXiv:0706.2334 [hep-ph]}}.

\bibitem{cteq6m}
D.~Stump et al.,  JHEP {\bf 10} (2003)  046,
\href{http://arxiv.org/abs/hep-ph/0303013}{{\tt arXiv:hep-ph/0303013}}.

\bibitem{Bozzi:2011en}
G.~Bozzi, F.~Campanario, M.~Rauch, and D.~Zeppenfeld,  Phys.Rev. {\bf D84}
  (2011)  074028,
\href{http://arxiv.org/abs/1107.3149}{{\tt arXiv:1107.3149 [hep-ph]}}.

\bibitem{Bozzi:2011wwa}
G.~Bozzi, F.~Campanario, M.~Rauch, and D.~Zeppenfeld,
  \href{http://dx.doi.org/10.1103/PhysRevD.83.114035}{Phys. Rev. {\bf D83}
  (2011)  114035},
\href{http://arxiv.org/abs/1103.4613}{{\tt arXiv:1103.4613 [hep-ph]}}.

\bibitem{geant4}
{GEANT4} Collaboration, S.~Agostinelli et al.,
\href{http://dx.doi.org/10.1016/S0168-9002(03)01368-8}{Nucl. Instrum. Meth.
  {\bf A506} (2003)  250}.

\bibitem{Aad:2010wq}
{ATLAS} Collaboration,
  \href{http://dx.doi.org/10.1140/epjc/s10052-010-1429-9}{Eur. Phys. J. {\bf
  C70} (2010)  823},
\href{http://arxiv.org/abs/1005.4568}{{\tt arXiv:1005.4568 [physics.ins-det]}}.

\bibitem{Aad:2008zzm}
{ATLAS Collaboration},
\href{http://dx.doi.org/10.1088/1748-0221/3/08/S08003}{JINST {\bf 3} (2008)
  S08003}.

\bibitem{Aad:2010sp}
{ATLAS Collaboration},
  \href{http://dx.doi.org/10.1103/PhysRevD.83.052005}{Phys. Rev. {\bf D83}
  (2011)  052005},
\href{http://arxiv.org/abs/1012.4389}{{\tt arXiv:1012.4389 [hep-ex]}}.

\bibitem{Aad:2011mk}
{ATLAS Collaboration},  Eur. Phys. J. {\bf C72} (2012)  1909,
\href{http://arxiv.org/abs/1110.3174}{{\tt arXiv:1110.3174 [hep-ex]}}.

\bibitem{Aad:2012re}
{ATLAS Collaboration},  Eur. Phys. J. {\bf C72} (2012)  1844,
\href{http://arxiv.org/abs/1108.5602}{{\tt arXiv:1108.5602 [hep-ex]}}.

\bibitem{Cacciari:2008gp}
M.~Cacciari, G.~Salam, and G.~Soyez,  JHEP {\bf 04} (2008)  063,
  \href{http://arxiv.org/abs/0802.1189}{{\tt arXiv:0802.1189}}.

\bibitem{Aad:2011he}
{ATLAS Collaboration}, {\em {Jet Energy Measurement with the ATLAS Detector
  in Proton-Proton Collisions at $\sqrt{s} = 7$ TeV}\/},
  \href{http://arxiv.org/abs/1112.6426}{{\tt arXiv:1112.6426 [hep-ex]}}.
  Submitted to Eur. Phys. J. C.

\bibitem{lumi2011}
{ATLAS Collaboration}, {\em {Luminosity Determination in $pp$ Collisions at
  $\sqrt{s} = 7$ TeV using the ATLAS Detector in 2011}\/},
  {ATLAS-CONF-2011-116}.
\newblock \url{http://cdsweb.cern.ch/record/1376384}.

\bibitem{Aad:2011dr}
{ATLAS Collaboration},
  \href{http://dx.doi.org/10.1140/epjc/s10052-011-1630-5}{Eur. Phys. J. {\bf
  C71} (2011)  1630},
\href{http://arxiv.org/abs/1101.2185}{{\tt arXiv:1101.2185 [hep-ex]}}.

\bibitem{Chou:2011}
G.~Choudalakis, {\em {On Hypothesis Testing, Trials Factor, Hypertests and
  the BumpHunter}\/},
\href{http://arxiv.org/abs/1101.0390}{{\tt arXiv:1101.0390 [physics.data-an]}}.

\bibitem{Aad:2011kz}
{ATLAS} Collaboration,
  \href{http://dx.doi.org/10.1140/epjc/s10052-011-1744-9}{Eur. Phys. J. {\bf
  C71} (2011)  1744}, \href{http://arxiv.org/abs/1107.0561}{{\tt
  arXiv:1107.0561 [hep-ex]}}.

\bibitem{Aad:2012xs}
{ATLAS Collaboration},
  \href{http://dx.doi.org/10.1140/epjc/s10052-011-1849-1}{Eur. Phys. J. {\bf
  C72} (2012)  1849},
\href{http://arxiv.org/abs/1110.1530}{{\tt arXiv:1110.1530 [hep-ex]}}.

\bibitem{MSTW2008}
A.~Martin, W.~Stirling, R.~Thorne, and G.~Watt,
  \href{http://dx.doi.org/10.1140/epjc/s10052-009-1072-5}{Eur. Phys. J. {\bf
  C63} (2008)  189},
\href{http://arxiv.org/abs/0901.0002}{{\tt arXiv:0901.0002 [hep-ph]}}.

\bibitem{Read:2002hq}
A.~L. Read,  \href{http://dx.doi.org/10.1088/0954-3899/28/10/313}{J. Phys. {\bf
  G28} (2002)  2693}.

\end{thebibliography}\endgroup
